\documentclass[twocolumn,preprintnumbers,amsmath,amssymb,prl,showkeys,showpacs]{revtex4}

\usepackage{graphicx}
\usepackage{amsfonts}
\usepackage{latexsym}

\begin{document}

\title{Open membranes, ribbons and deformed Schild strings}

\author{David S. Berman}
\email{D.S.Berman@qmul.ac.uk} 
\affiliation{ Department of Physics, \\Queen Mary College University of 
London, Mile End Road,\\ London E1 4NS, England }

\author{Boris Pioline}
\email{pioline@lpthe.jussieu.fr}
\affiliation{LPTHE, Universit\'es Paris VI et VII, 4 pl Jussieu, \\
75252 Paris cedex 05, France}

\preprint{LPTHE-04-06, QMUL-PH-04-02 }

\begin{abstract}
We analyze open membranes immersed in a magnetic three-form
field-strength $C$. While cylindrical membranes in the absence of
$C$ behave like tensionless strings, when the $C$ flux is present the
strings polarize into thin membrane ribbons, locally orthogonal to the
momentum density, thus providing the strings with an effective
tension. The effective dynamics of the ribbons can be described by a
simple deformation of the Schild action for null
strings. Interactions become non-local due to the polarization, and
lead to a deformation of the string field theory, whereby string
vertices receive a phase factor proportional to the volume swept out
by the ribbons. In a particular limit, this reduces to the
non-commutative loop space found previously.
\end{abstract}



\maketitle

\def    \bea    {\begin{eqnarray}}
\def    \eea    {\end{eqnarray}}
\def    \beq    {\begin{equation}} 
\def    \eeq    {\end{equation}}
\def    \be     {\begin{equation}} 
\def    \ee     {\end{equation}}
\def    \lf     {\left (} 
\def    \rt     {\right )}
\def    \a      {\alpha} 
\def    \lm     {\lambda}
\def    \D      {\Delta} 
\def    \r      {\rho}
\def    \th     {\theta} 
\def    \rg     {\sqrt{g}} 
\def    \Slash  {\, \! \! \! \!}  
\def    \comma  {\; , \; \;} 
\def    \pl     {\partial} 
\def    \del    {\nabla}
\def \a {\alpha}
\def \b {\beta}
\def \g {\gamma}
\def \pa {\partial}
\def \ap {}
\def \eps {\epsilon}
\def \diag {\mbox{diag}}

Open strings in large magnetic backgrounds at low energy are best
described as excitations of a non-commutative Yang-Mills theory. While
interesting questions remain unanswered as to their precise dynamics and
observables, non-commutative gauge theories are well defined field
theories, albeit with an infinite series of higher derivative
interactions. The Moyal star product entering their Lagrangian
is a simple consequence of the nature of the microscopic
degrees of freedom, which behave like non-relativistic
elastic dipoles in a strong magnetic field \cite{Bigatti:1999iz}.

In contrast, open membranes, whether in non-trivial 3-form backgrounds or in
vacuo, are much more elusive objects. They are believed to be the
appropriate degrees of freedom of the 5-brane on which they are
required to end, yet the precise way in which the self-dual dynamics
of the two-form gauge potential (and other members of the
$(2,0)$-supersymmetric multiplet on the 5-brane) arises by
quantization of the open membrane is completely mysterious. This is
not to mention of course the case of a stack of $N$ 5-branes, for which
the two-form analogue of non-Abelian Yang-Mills is not known. Indeed,
there are even reasons to doubt that a field theory description may be
appropriate, since membranes ending on two coinciding 5-branes
presumably behave like tensionless strings, with an infinite tower of
states and no tunable coupling (see however \cite{Henningson} for
recent progress). 

Nevertheless, there are reasons to believe that membranes in
magnetic backgrounds should be more tractable. For one, we do
not expect that switching on a magnetic or electric  background
will change the spectrum: the (2,0) multiplet should still 
describe open membranes ending on a single 5-brane with a $C$ field.
Second, in a large 3-form field strength the membrane dynamics 
should be dominated by the boundary coupling
\beq 
\int_{\pa M_2} ~C_{\mu \nu \rho} ~X^{\mu} ~ d X^{\nu} 
\wedge dX^{\rho} 
\label{baction}
\eeq 
which should be easier to quantize than the Nambu-Goto part of
the membrane action: indeed, in   the string theory   case,
quantizing the first-order quantum mechanics on the boundary 
of the  open string is the most direct path to the  Moyal product.
Steps in this direction have been reviewed in \cite{boris},
and have lead  to a heuristic proposal for the leading deformation
of Abelian two-form dynamics, compatible with invariance under
volume-preserving diffeomorphisms \cite{boris}.

The purpose of this paper is to pursue the analysis of open 
membranes in a large magnetic $C$ field at a purely classical level,
building on earlier  work \cite{boris,Matsuo:2001fh}.

We first observe that the relevant kinematical degrees of freedom are membranes
with two boundaries, and cylindrical topology, which in the 
absence of $C$ would behave as tensionless strings. Our main finding is
that the magnetic 3-form polarizes these strings into thin
ribbons, orthogonal to the local momentum density. The ``open membrane
theta parameter'', first introduced by indirect methods in
\cite{Berman:2001rk}, is now physically interpreted as the polarizability
of these ribbons. In this process, the originally tensionless strings
gain inertia and can be described as non-relativistic tensionful
strings, albeit with a non-standard worldsheet action. Just as with electric
dipoles in a magnetic field, this polarization induces non-local interactions,
which can be summarized by a deformation of the closed string field
theory, analogous to the non-commutative deformation. In a particular
gauge, we recover the non-commutative string found in 
\cite{bergshoeff,Kawamoto:2000zt}.

A possible concern with this line of reasoning is the 
fact, exploited in the OM proposal \cite{OM}, 
that a magnetic three-form field strength,
$H_{123}$, cannot be  large unless the dual electric component 
$H_{045}$ approaches the critical electric field value.
While membranes with a single boundary in this limit
tend to grow in the (045) directions until they break, 
cylindrical membranes remain tensionless in the 
$(045)$ directions. The turning on of the $H_{045}$ components therefore will not qualitatively affect our conclusions.

Finally, it would be useful to make contact with more
more formal approaches such as
\cite{Park:2000au,Hofman:2001zt,Hofman:2002jz}, 
which have focussed on the deformation of the $L_{\infty}$ algebra
underlying the closed string field theory, in the small $C$ limit.

\section*{The non-commutative string, revisited}

As a starting point, let us recall some basic features of the dynamics of an
open string in a magnetic field as originally described in
\cite{Bigatti:1999iz} with some additional details to aid the
generalisation to the membrane. In the process, we shall gain
understanding of the open string ``metric'' and ``non-commutativity
parameter'' which we will be able to extend to the membrane case.

\enlargethispage{7mm}

Open strings end on D-branes.  When only one end is 
immersed in a magnetic field, the string behaves like a charged
particle and being trapped on Larmor orbit, takes little part in the transport
properties of the system. In contrast, when both
of the ends are immersed in  the same magnetic  field, the
string is globally neutral and  behaves like  an electric dipole.
When moving at a velocity $\vec v$,  the magnetic Lorentz force 
$\vec F_l=\pm e \vec v
\wedge \vec B$  exerted at the end must cancel the elastic force
$\vec F_e=  \pm k \vec \Delta$, where $k$ is  the elasticity constant,
leading to a  polarisation $\Delta = e \vec v \wedge \vec B  /k$ transverse
to  the direction of motion.  A useful analogy is  that of  vortices
in two-dimensional fluid  dynamics: by the Magnus force, two vortices
of opposite vorticity are able to propagate forward, the velocity
field of one carrying the other along.

To see how  this  non-relativistic description emerges from the
usual relativistic funcamental string, recall that the electromagnetic
coupling imposes the boundary condition
\be
\pa_\sigma X^i +  B_{ij} ~ \pa_\tau X^j = 0\ ,\quad
~\mbox{at}~ \sigma=0,\pi   \, .
\ee
This can be solved along with the bulk equation of motion
$(\pa_\sigma^2- \pa_\tau^2)X^\mu=0$ to give the zero-mode solution,
\be
X^i=  p_0^i \tau +  B_{ij} ~ p_0^j \sigma 
\ee
(we set the string tension to $1$, and assume that
the target space metric is that of Minkowski space; indices
are raised or lowered with the Kroneker symbol $\delta_{ij}$). 
From this expression, it is apparent that the string is stretched
into a dipole of length
\be
\Delta_i= B_{ij} ~ p_0^j \, .
\ee
The canonical linear momentum is related to $p_0$ by
\be
P^i = 
\left( \pa_\tau X^i -  B_{ij} \pa_\sigma X^j \right)
= (1+ B^2)  ~p_0^i
\ee
so indeed the dipole is stretched proportionally to its
momentum $P^i$. This gives an elongation
\be
\label{dp}
\Delta^i = \Theta^{ij} P_j\ ,\qquad 
\Theta = \frac{B}{1+B^2}    \, . 
\ee
The ``open string non-commutativity parameter'' $\Theta^{ij}$, 
introduced in \cite{SW}, can thus be viewed as  the dynamic polarizability
of the open string dipole in a magnetic field. As 
argued  
in \cite{Bigatti:1999iz},
the fact that open string dipoles interact via their end points 
implies non-local interactions in the effective field theory,
e.g. for two point interactions,
\bea
{\cal L} &\sim& \int d^n  x~ 
\phi_1 \left( x + \frac12 \Theta^{ij} P_j  \right) 
\phi_2 \left( x - \frac12 \Theta^{ij} P_j  \right) \\
&=& \int d^n x  ~ \phi_1 * \phi_2
\eea
which is precisely the effect of the Moyal star product.
More generally, $n$-point vertex are pick up a phase
proportional to the area of the polygon formed by the
incoming dipoles.

To see how the ``open string metric'' arises, let us  compute the
total energy of the string. Using that $X^0 = \alpha' E \tau$, 
and that the worldsheet Hamiltonian,
\be
H = \int (\pa_\tau X)^2 + (\pa_\sigma X)^2  =
 (1+ B^2)  p_0^2 - E^2  + N
\ee
should vanish on physical states ( $N$ is the contribution
of the excited levels of the string) the energy is given by the following dispersion relation
\be
\label{enf}
E = \sqrt{ m^2 +  G^{ij} P_i P_j}\ ,\quad
G_{ij}  = (G^{ij})^{-1} = (1+  B^2) \delta^{ij}
\ee
where $m^2 = N$. The effective metric $G_{ij}$
governing the dependence of the energy on the momentum
is the ``open  string  metric'' as discussed in \cite{SW}.

One can now check that the balance between  the Lorentz
and tensive forces is satisfied.  The velocity is $v^i = p_0^i / E$, 
hence the Lorentz force is given by:
\be
F_l^i = B_{ij} v^j  = \frac{1}{E} \Delta^i \, .
\ee
On the other hand, expressing the momentum $P$ as a function of 
the elongation $\Delta$  into  the energy \eqref{enf}, one derives
\be
F_t^i = \frac{\pa E}{\pa \Delta^i}
= \frac{1+ B^2}{B^2} \frac{\Delta^i}{E}  \, .
\ee
This indeed cancels the Lorentz force  $F_l$ in the limit of large 
magnetic field $B$, validating the assumption that the string
is in its ground  state.

It is instructive to redo this computation without fixing the 
conformal gauge (as this  is not available in the membrane case). 
The equations of motion and boundary conditions for $X^\mu$ read 
\bea
\pa_\alpha  (\sqrt{\gamma}  \gamma^{\alpha\beta} \pa_\beta X^\mu) &=& 0\\
\sqrt{\gamma} \gamma^{\sigma\sigma} \pa_\sigma X^i +  B_{ij} 
\pa_\tau X^j &=& 0
~~\mbox{at}~~\sigma=0,\pi
\eea
where $\gamma_{ab}$ has to  be  equal to the induced metric $\pa_{\a}X^\mu
\pa_{\b}X_\mu$, up to a conformal factor. The zero-mode ansatz
\be
X^i =  p_0^i \tau + \Delta^i \sigma  \ ,\qquad
\gamma_{\a\b} = \diag(m^2,\Delta^2)
\ee
where $m^2= E^2 - p_0^2$,  is thus a solution if
\be
\frac{\Delta^i}{|\Delta|} = \ap B_{ij}~ \frac{p_0^j}{m} \, .
\ee
While this equation does  not specify  the length of $\vec \Delta$,
upon substituting $p_0$ by the physical momentum 
\be
P_i = 
\sqrt{\g} \g^{\tau\tau} \pa_\tau X^i
- B_{ij}  \pa_\sigma X^j = \frac{|\Delta|}{m} ~ (1+B^2)~
p_0^i
\ee
it becomes equivalent  to  the relation \eqref{dp} above.

\enlargethispage{7mm}

\section*{From strings to ribbons}
We now come to the case of non-commutative membranes in a large
$C_{123}$ magnetic flux. Our first  observation is that,  in order
to contribute to the transport properties of the (2,0) theory, the
membrane should have at least two boundaries on the 5-brane
with  non-vanishing flux. This is because, as  noticed in  \cite{boris},
for a  single boundary  the equations of motion of 
the boundary string
\be
C_{ijk} ~ \pa_\sigma X^j \wedge \pa_\tau X^k = 0 
\ee
imply that $X^i$ has to depend on the  worldsheet coordinates
through one function $f(\tau,\sigma)$ only: the boundary of the
membrane therefore  spans a  static string in the $(X^1,X^2,X^3)$
plane. A useful analogy is that of a vortex
line in a three-dimensional  fluid: just as in two-dimensions, 
the transverse motion  of a vortex line  is
effectively confined by the rotational motion of the fluid itself,
on Landau-like orbits. In addition, there exist soft  modes
propagating along the vortex line  known as Kelvin modes. In fact,
as noticed  in  \cite{Matsuo:2001fh,boris}, 
the boundary coupling \eqref{baction} is precisely
the one describing  the Magnus effect in fluid hydrodynamics.
Of course, in contrast to the  two-dimensional case, the  total
vorticity is not conserved  and  a vortex  line may slowly shrink
and disappear, just as the membrane boundary may shrink to a point
and leave the five-brane, under  the effect of its tension.

On the other hand, membranes with two boundaries have no overall
charge and therefore can propagate freely:  this is the analogue of
configurations of vortex anti-vortex lines in hydrodynamics.  In the
absence of a $C$ field (and with no Higgs vev),  the two boundaries
lie on top of each other, leading to an effectively tensionless string,
the tentative fundamental degrees of freedom
of the (2,0) theory.  In the presence of a magnetic field however, it
is easy to see  that these tensionless strings polarize into thin
ribbons, whose width is proportional to the local momentum density.
Indeed, the canonical momentum on the membrane, neglecting the
contribution of the Nambu-Goto term, is
\be
P^i =C_{ijk} ~\partial_\sigma X^j \partial_\rho X^k
\ee
where $\sigma$ is the coordinate along the boundary string, and
$\rho$ the coordinate normal to it. The ribbon thus grows as
\be
\Delta^i \sim \partial_\rho X^i= 
\frac{1}{C |\pa_\sigma X|^2} \eps_{ijk} P^j \pa_\sigma X^k 
\ee
where we retain in $\Delta$ only the component orthogonal to 
$\sigma$ (the parallel component could be reabsorbed by
a diffeomorphism on the membrane worldvolume).

In order to study more precisely this polarization, let us consider a
simple classical solution corresponding to an infinite strip of width
$\Delta$ moving at a constant velocity $v$ transverse to it: We thus
consider the classical solution 
\be 
\label{zmm}
X^i = p_0^i \tau + u^i \sigma + \Delta^i \rho \; \; . 
\ee 
The boundary condition 
\be  \sqrt{\gamma}
\gamma^{\rho\rho} \pa_\rho X^i - C_{ijk} \pa_\sigma X^j \pa_\tau X^k
= 0 
\ee 
with induced metric $\gamma=\diag(m^2,|u|^2,|\Delta|^2)$, implies
that the direction of the polarization vector is orthogonal to the plane
formed by the tangent vector to the string $\vec u=\partial_\sigma \vec X$
and the local velocity $\vec p_0 = \partial_\tau \vec X$, 
\be
\frac{\vec \Delta}{|\Delta|} = C \frac{\vec u}{|u|} \wedge \frac{\vec
p_0}{m} \, .
\ee 
Calculating the local canonical momentum
\bea 
\label{cm}
P^i &=& 
\sqrt{\gamma} \gamma^{\tau\tau} \pa_\tau X^i - C_{ijk} ~ \pa_\sigma
X^j ~ \pa_\rho X^k \\ &=&
\frac{|\vec u|~|\vec\Delta|}{m(1+C^2)} ~\left[ ( 1+ C ^2) p_0^i -
C^2 \frac{\vec u \cdot \vec p_0}{|\vec u|^2}  u^i \right] 
\eea 
one may express the local velocity in terms of $P^i$,
\be
\label{poi}
p_0^i = \frac{m}{|\vec u|~|\vec\Delta|(1+C^2)}
\left[ P_i + C^2 \frac{\vec u \cdot \vec P}{|\vec u|^2} u^i \right] 
\ee
and obtain the relationship between the membrane polarization and 
canonical momentum,
\be \vec \Delta = \Theta \frac{\vec
u}{|u|^2} \wedge \vec P\ \ , \qquad \Theta = \frac{C}{1+C^2} \, .
\label{th}
\ee
For convenience, we will use the gauge $m=|\vec u|~|\vec \Delta|$
from now on.

We thus recover the ``open membrane non-commutativity parameter'' $\Theta$,
defined in \cite{Berman:2001rk}. In this work, 
the ``open membrane non-commutativity
parameter'' and ``open metric'' were determined by studying
the physics of five-branes probing
supergravity duals with $C$-flux longitudinal to the probe brane world
volume. We now understand this result classically as the 
the polarizability of open membranes in a $C$-field. 
We shall return to the issue of the ``open membrane metric'' shortly.

\enlargethispage{7mm}

\section*{An effective Schild action for string ribbons}

Membranes are notoriously difficult to quantize. Since the
effect of the magnetic background is to polarize the boundary strings,
hence given a  non-zero tension, it is interesting to ask if one
can write down an effective string theory that may be more tractable.
For this,  let us start with the light-cone formulation \cite{hoppe} 
of the membrane, with Hamiltonian
\be
\label{hop}
P^- = \int d\sigma d\rho ~\frac{1}{2P^+}
\left[ (p_0^i)^2  +  g \right]
\ee
where $g$ is the determinant of the spatial metric, hence the square
of the area element (and the membrane tension is set to 1). 
In this gauge, one should enforce the constraint 
\be
\pa_\sigma X^i \pa_\rho \pa_\tau X^i - 
\pa_\rho X^i \pa_\sigma \pa_\tau X^i =0
\ee
which is trivailly satisfied on zero-mode configurations \eqref{zmm}.
For a thin ribbon of width $\vec\Delta$ given
by \eqref{th}, the square of the area element is
\be
g=| \vec u \wedge \vec \Delta|^2 =
\frac{C^2}{(1+C^2)^2} \left[ \vec P^2 - \frac{(\vec u \cdot \vec P)^2}
{|u|^2} \right]
\ee
On the other hand, using \eqref{poi}, the kinetic energy may be written as
\be
(\vec p_0)^2 = \frac{1}{(1+C^2)^2} 
\left[ \vec P^2 + C^2 (C^2+2) \frac{(\vec u \cdot \vec P)^2}{|u|^2} \right]
\ee
The total Hamiltonian thus takes the form
\be
\label{h1}
P^- = \int d\sigma \frac{1}{2P^+ (1+C^2)} 
\left[ P^2 + C^2 \frac{(\vec P \cdot 
\pa_\sigma \vec X)^2}{| \pa_\sigma X^i|^2} \right]
\ee
From this expression, specifying to a gauge choice where $\vec P$
and $\pa_\sigma \vec X$ are orthogonal, we see that the effective
metric in the transverse directions is rescaled by a factor of
$(1+C^2)$,
\be
G_{ij} = \left[ 1 + C^2 \right] \delta_{ij}
\ee
This agrees with the membrane metric found from very different considerations
in \cite{Berman:2001rk,VanderSchaar:2001ay}, 
up to the conformal factor  $Z=(1-\sqrt{1-1/K^2})^{1/3}$ 
with $K=\sqrt{1+C^2}$. It should however be noted that the
evidence for this conformal factor is rather indirect, and its
non-analyticity  $Z\sim 1-|C|/3 + O(C^2)$ at weak $C$ is not
understood. It is also possible that quantum corrections 
on the membrane  world-volume may correct our classical result.

Finally, we may perform a Legendre transform on $P_i$
to find the Lagrangian density of the ribbon,
\be
\label{lag}
{\cal L}=\int d\sigma ~\frac12 (\pa_\tau X^i)^2 
+ \frac{C^2}{2| \pa_\sigma X^i|^2} \sum_{i,j} 
\{ X^i,X^j \}^2
\ee
where we have defined the Poisson bracket on the Lorentzian string 
worldsheet\footnote{This should not be confused with the Poisson bracket
formulation of the membrane, which refers to the two {\it spatial} directions
of the membrane world-volume.} as $\{A,B\}=\pa_\sigma A \pa_\tau B - \pa_\sigma B \pa_\tau A$. Note that 
the relative sign between the two terms in \eqref{lag} is consistent
with the fact that they both contribute to kinetic energy. For vanishing
$C$, \eqref{lag} reduces to the Lagrangian for a tensionless string,
as expected. While we have mostly worked at the level of zero-modes,
it is easy to see that \eqref{lag} remains correct for arbitrary
profiles $X^i(\tau,\sigma)$, as long as the dependence on 
membrane coordinate $\rho$ is fixed by Eqs. \eqref{zmm}, \eqref{th}.

After fixing the invariance
of the Lagrangian \eqref{lag} under general reparameterizations of $\sigma$
by choosing $| \pa_\sigma X^i|=1$, we recognize in the second term the Schild 
action, which provides (in the case of a  Lorentzian target-space)
a unified description of both tensile and tensionless strings, 
depending on the chosen value for the conserved quantity 
$\omega=\{ X^i , X^j \}^2$~\cite{Schild:vq}. This term dominates over the first
in the limit of large $C$ field. For any finite value however, $\omega$ is
not conserved, and the second term in \eqref{lag} can be interpreted as
the action for a non-relativistic string with tension proportional to $C$.

As usual, it is possible to give a regularization of this membrane
action, by replacing the Poisson bracket (now in light-cone directions
on the worldsheet) by commutators in a large $N$ matrix model.
One thus obtains a lower-dimensional analogue of the type IIB IKKT
matrix model \cite{Ishibashi:1996xs},
\be
P^- = \frac{1}{2P^+} \left(
[A_0,X^i]^2 
+  C^2  \sum_{i<j} [ X^i, X^j ] ^2 \right)\, .
\ee
It would be interesting to understand how the matrix regularization
distinguishes the Lorentzian worldsheet from more usual Euclidean one.
We leave the study of this
model and its supersymmetric  version for future  work.

\section*{Non-commutative string field theory}

Just like open strings, open membranes interact only when their
ends coincide. Since their  boundaries  are tensionless
closed strings which polarize into thin
ribbons in the presence of a strong $C$ field, one may expect
that the effect  of  the $C$  field can be encoded by a deformation
of the  string field theory describing the membranes boundaries.
Despite the fact that string  field theory  of closed  strings, 
not to mention tensionless ones, is a rather ill-defined subject,
it is natural to represent the string field as a functional 
in the space of loops. The effect of the polarization of
the ribbons can thus represented  by
\bea
\label{vstar}
V &\sim& \int [DX(i)] 
\Phi\left[ X^i- \frac12 \frac{\Theta}
{ |\pa_\sigma X|^2} \eps_{ijk} \pa_\sigma X^j 
\frac{\delta}{\delta X^k}\right]~ \nonumber\\ &&\times
\Phi\left[ X^i+ \frac12 \frac{\Theta}
{ |\pa_\sigma X|^2} \eps_{ijk} \pa_\sigma X^j 
\frac{\delta}{\delta X^k}\right]
\eea
where we represented  the momentum density $P_i$, canonically
conjugate to $X^i(\sigma)$, as a derivative operator in the space of loops.
Defining the operators
\be
\tilde X^i(\sigma) =  X^i- \frac12 \frac{\Theta}{ |\pa_\sigma X|^2}
\eps_{ijk} \pa_\sigma X^j \frac{\delta}{\delta X^k}
\ee
it is easy to reproduce the non-commutative loop space in the
``static'' gauge $X^3(\tau,\sigma)=\sigma$,
\be
[ \tilde X^1(\sigma), \tilde X^2(\sigma') ]
= \Theta \delta(\sigma-\sigma')
\ee
as proposed in \cite{bergshoeff,Kawamoto:2000zt}. 
The fact that the transverse fluctuations of a vortex line
are effectively confined by an harmonic potential is well
known in fluid dynamics. In the more covariant
gauge $|\pa_\sigma \vec X|=1$, one  obtains a
tensionless limit of the $SU(2)$ current algebra,
\be
[ \tilde X^i(\sigma), \tilde X^j(\sigma') ]
= \Theta \eps_{ijk} \pa_\sigma X^k \delta(\sigma-\sigma')  \, .
\ee
The same relations may be directly obtained by Dirac quantization
of the topological open membrane Lagrangian \eqref{baction}.

More generally, much as  in the non-commutative
case, this deformation amounts to multiplying the closed string
scattering amplitudes by a phase factor proportional to the
volume enclosed  by the ribbons as  they interact. It would
be very interesting to derive the deformation of the 
$(2,0)$ effective field theory from \eqref{vstar},  and possibly
verify the proposal in \cite{boris} motivated by the invariance
under volume preserving diffeomorphisms.

\enlargethispage{7mm}

{\it Acknowledgments}:  The authors would like to thank DAMTP in Cambridge 
and LPTHE in Paris for hospitality during part of this project.  B. P. is 
grateful to discussions with R. Gopakumar, C. Hofman, L. Motl and
S. Minwalla at several stages of this work. D.S.B
wishes to thank Cambridge University and Clare Hall College for continued
support and is funded by EPSRC grant GR/R75373/01.

\end{document}